\documentclass[10pt,preprintnumbers,showpacs,amsmath,amssymb,onecolumn]{revtex4}
\input epsf.sty
\usepackage{graphicx}
\usepackage{amssymb}
\usepackage{txfonts}
\usepackage{bm}% bold math
\usepackage{graphicx,amsmath,mathrsfs}
\begin{document}
\title{Collective Multipole excitations of exotic nuclei in relativistic
continuum random phase approximation\footnote{Supported by the
National Natural Science Foundation of China under Grant Nos
11175216, 11275018 and 11305270, and the National Basic Research
Program of China under Grant No 2013CB834404, and Science Planning
Project of Communication University of China (XNL1207)}}
\author{Yang Ding$^{a,b}$, Li-Gang Cao,$^{c,d}$ Ma Zhongyu$^{b}$\\
\noindent a. School of Science, Communication University of China, Beijing 100024\\
\noindent b. China Institute of Atomic Energy, PO Box 275(18), Beijing 102413\\
 \noindent c. Center of Theoretical Nuclear Physics, National Laboratory of Heavy Collision, Lanzhou 730000\\
 \noindent d. Institute of Modern Physics, Chinese Academy of Sciences, Lanzhou
730000
 }
\date{\today}
\begin{abstract}

The  isoscalar and isovector collective multipole excitations in
 %Ca isotopes
 exotic nuclei are studied in the framework of a fully self-consistent
relativistic continuum random phase approximation (RCRPA). In this
method the contribution of the continuum spectrum to nuclear
excitations is treated exactly by the single particle Green's
function. %In comparison with
Different from the cases in stable nuclei, there are strong
 low-energy excitations in neutron-rich nuclei
and proton-rich nuclei. The neutron or proton excess pushes the
centroid of the strength function to lower energies and increases
the fragmentation of the strength distribution. The effect of
treating the contribution of continuum exactly are also discussed.
%and it is found that the coupling between the bound states and the
%continuum becomes important in neutron-rich nuclei and proton-rich
%nuclei.


\end{abstract}

  %Nuclear matter}}
\maketitle

\section{Introduction}

The structure of nuclei far from $\beta$-stable  line is an exciting
research field since a number of new phenomena are expected or have
been observed in neutron-rich nuclei and proton-rich nuclei.
Currently, physicists are interested in the study of the effect of
neutron or proton excess on various collective excitations. As a
result, the multipole response of %unstable and
exotic nuclei, such as neutron-rich nuclei and proton-rich nuclei,
becomes  a rapidly growing research field\cite{Paar}. Collective
excitations can be studied by the self-consistent relativistic
random phase approximation (RRPA) built on the relativistic mean
field (RMF) ground state\cite{Ma01,Ma02,Ring01,Pie01}. However, in
most previously random phase approximation calculations the
contribution of the continuum %near the threshold
might not be treated properly since the nucleon states in the
continuum are discretized by a basis expansion or by setting a box
approximation. The coupling between the bound states and the
continuum becomes important since the Fermi surface is close to the
particle continuum in %unstable and
exotic nuclei\cite{Cao04}. As a result, when one works on the
properties of nuclei far from the $\beta$-stable line, it is
required to consider the contribution of the continuum rigorously.

The fully self-consistent relativistic continuum random phase
approximation (RCRPA) has been constructed in the momentum
representation\cite{YangDing2009,YangDing20101,YangDing20102,YangDing2010prc}.
In this method the contribution of the continuum spectrum to nuclear
excitations is treated exactly by the single particle Green's
function. In this work, in order to clarify  the effect of neutron
or proton excess on  collective excitations, the RCRPA method is
used to study the isoscalar and isovector multipole collective
excitations in neutron-rich , proton-rich and $\beta$-stable nuclei.
Here we want to explore the effect of neutron (proton) excess on the
strength distribution of multipole collective excitations, for
simplicity, we choose some sub-closed shell nuclei and magic nuclei
for our motivation, such as $^{34}$Ca, $^{40}$Ca, $^{48}$Ca ,
$^{60}$Ca , $^{16}$O , $^{28}$O , $^{100}$Sn and $^{132}$Sn, the
pairing in sub-closed shell nuclei is not included in the present
study.

The outline of this paper is as follows. The fully self-consistent
RCRPA  is introduced in Sec.II. In Sec.III we discuss collective
multipole excitations of exotic nuclei in RCRPA. Finally we give a
brief summary in Sec.IV.

\section{Fully Consistent  RELATIVISTIC CONTINUUM   RANDOM  PHASE APPROXIMATION  }

  We start from the single particle Green's function
which  is defined by:
\begin{eqnarray}
G({\bf r,r'};E)=\sum\limits_{b = h,A,\bar {\alpha }} {\frac{f_b({\bf
r})\bar{f}_b({\bf r'})}{E_b - E \mp i\eta }} ~,\label{eq4}
\end{eqnarray}
The Green's function can be decomposed into radial functions
$g_{ij}(r,r')$ and spin-spherical harmonics, then we can get the
radial equation:
\begin{equation}
\left( {{\begin{array}{*{20}c}
 { - M^\ast - V + E} \hfill & d/dr - \kappa /r \hfill \\
 d/dr + \kappa /r \hfill & { - M^\ast + V - E} \hfill \\
\end{array} }} \right)\left( {{\begin{array}{*{20}c}
 {g_{11}^\kappa(r,r') } \hfill & {g_{12}^\kappa(r,r') } \hfill \\
 {g_{21}^\kappa(r,r') } \hfill & {g_{22}^\kappa(r,r') } \hfill \\
\end{array} }} \right)%\nonumber\\
= \delta (r - r')\left( {{\begin{array}{*{20}c}
 { - 1} \hfill & 0 \hfill \\
 0 \hfill & 1 \hfill \\
\end{array} }} \right)~.\label{eq8}
\end{equation}
 The regular and irregular solutions of the radial equation are
\begin{equation}
u^\kappa(r) = \left( {{\begin{array}{*{20}c}
 {u_1^\kappa(r) } \hfill \\
 {u_2^\kappa(r) } \hfill \\
\end{array} }} \right)~,~~~~~~~~~
v^\kappa(r) = \left( {{\begin{array}{*{20}c}
 {v_1^\kappa(r) } \hfill \\
 {v_2^\kappa(r) } \hfill \\
\end{array} }} \right) ~.\label{eq9}
\end{equation}
In terms of $u^\kappa$ and $v^\kappa$,the radial Green's functions
are given  by
\begin{eqnarray}
&& g_{ij}^\kappa (r,r')
%\nonumber \\&&
= \frac{1}{\Delta ^\kappa }\left\{ {{\begin{array}{*{20}c}
 {u_i^\kappa (r)v_j^\kappa (r'){\begin{array}{*{20}c}
 , \hfill & {r \le r'} \hfill \\
\end{array} }} \hfill \\
 {v_i^\kappa (r)u_j^\kappa (r'){\begin{array}{*{20}c}
 , \hfill & {r > r'} \hfill \\
\end{array} }} \hfill \\
\end{array} }} \right.~~{\begin{array}{*{20}c}\hfill & {i,j = 1,2} \hfill \\
\end{array} }~.\label{eq10}
\end{eqnarray}
where $\Delta ^\kappa = v_1^\kappa u_2^\kappa - u_1^\kappa
v_2^\kappa$  is the Wronskian determinant.

 The response function of a quantum system to
an external field is given by the imaginary part of the retarded
polarization operator,
\begin{equation}
R(P,P;E)=\frac 1\pi Im\Pi ^R(P,P;{\bf k},{\bf %
k^{\prime }};E)|_{{\bf k=k'}=0}~,   \label{eq1}
\end{equation}
where $P=\gamma _{0}r^{J}Y_{JM}(\hat{r})$ for the isoscalar electric
multipole excitation  and multiplied by $\tau_{3}$ for isovector
 excitation.
 The RCRPA polarization operator is obtained by
solving the Bethe-Salpeter equation,
%\begin{widetext}
\begin{eqnarray}
&\Pi(P,P;{\bf k},{\bf k^{\prime }},E)=&\Pi _0(P,P;{\bf k},{\bf
k^{\prime }} ,E)-\sum_ig_i^2\int d^3k_1d^3k_2\Pi _0(P,\Gamma ^i;{\bf
k},{\bf k}_1,E) D_i({\bf k}_1,{\bf k}_2,E)\Pi (\Gamma _i,P;{\bf
k}_2,{\bf k^{\prime }},E)~. \label{eq2}
\end{eqnarray}
%\end{widetext}
we can obtain  the unperturbed retarded polarization operator $\Pi
_0$ in the RMF ground state:
\begin{eqnarray}
 \Pi _0^{\rm RMF} (P,Q;{\bf r,r'};E)
& %=& \sum\limits_h {\left[ {\bar {f}_h ({\bf r})P^{+} \sum\limits_{b
%= h',A,\bar {\alpha }} {\frac{f_b ({\bf r})\bar {f}_b ({\bf
%r'})}{E_b - (E_{h'} + E) - i\eta }Qf_h ({\bf r'})} } \right.} +
%\left. {\bar {f}_h ({\bf r'})Q\sum\limits_{b = h',A,\bar {\alpha }}
%{\frac{f_b ({\bf r'})\bar {f}_b ({\bf r})}{E_b - (E_{h'} - E) +
%i\eta }P^{+} f_h ({\bf r})} } \right] \nonumber\\
=& \sum\limits_h {\left[ {{\bar {f}_h ({\bf r})P^{\dag} G({\bf
r,r'}; \omega^+ ) Qf_h ({\bf r'})} } \right.} + \left. { {\bar {f}_h
({\bf r'})Q G({\bf r',r}; \omega^- ) P^{\dag} f_h ({\bf r})}
 } \right] ~,\label{eq5}
\end{eqnarray}
where $\omega^\pm = E_{h} \pm E \pm i\eta$, %and $\omega^- = E_a - E -i\eta$,
%$E_{h}$ are the eigenvalue of hole states, $E$ is the excitation
%energy.

\begin{figure}[t]
\begin{center}
   \resizebox{20pc}{!}{\includegraphics{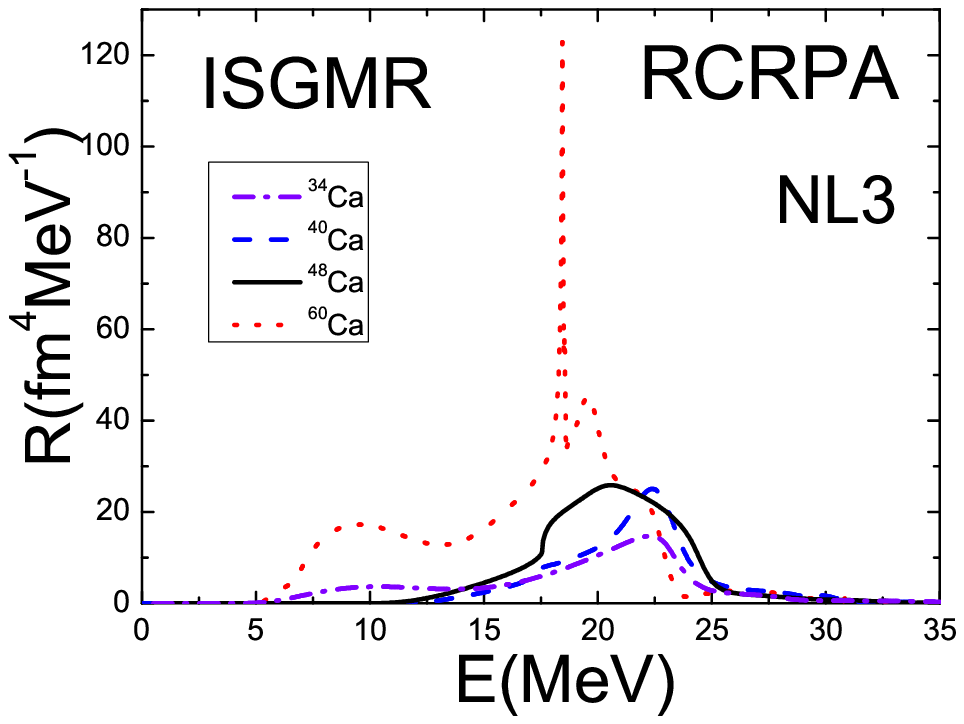}}
   \resizebox{20pc}{!}{\includegraphics{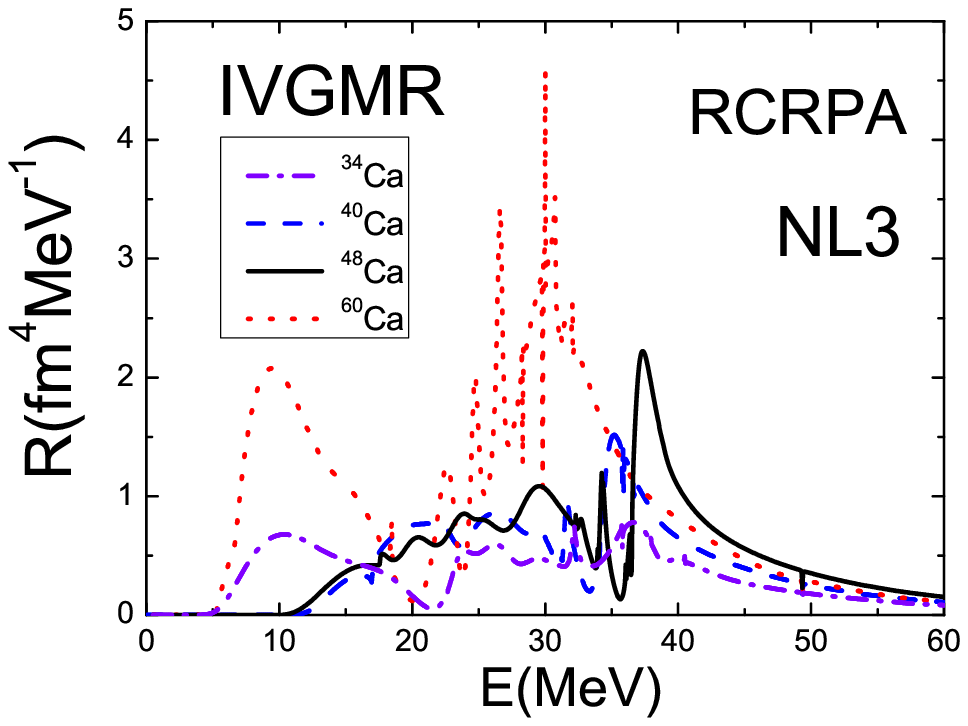}}
  \resizebox{20pc}{!}{\includegraphics{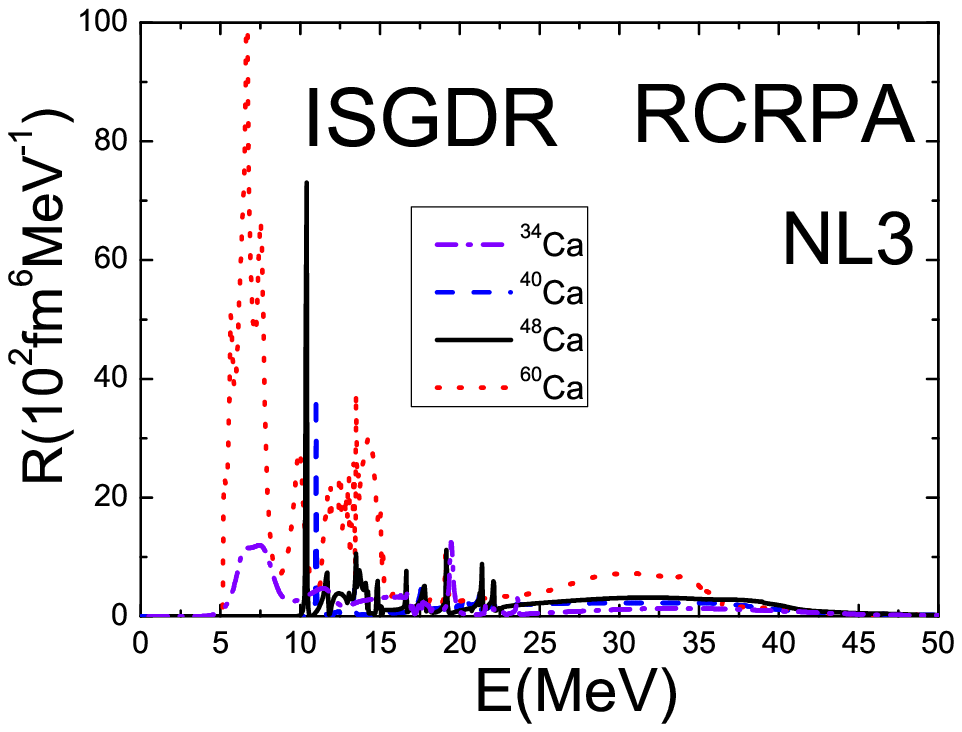}}
 \resizebox{20pc}{!}{\includegraphics{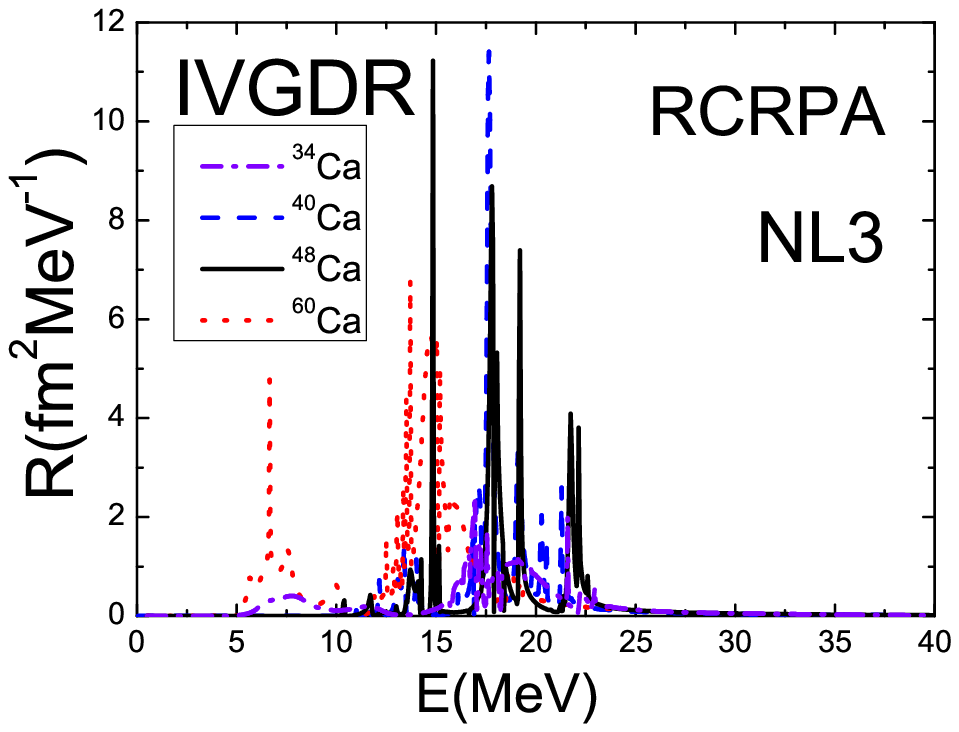}}
   \resizebox{20pc}{!}{\includegraphics{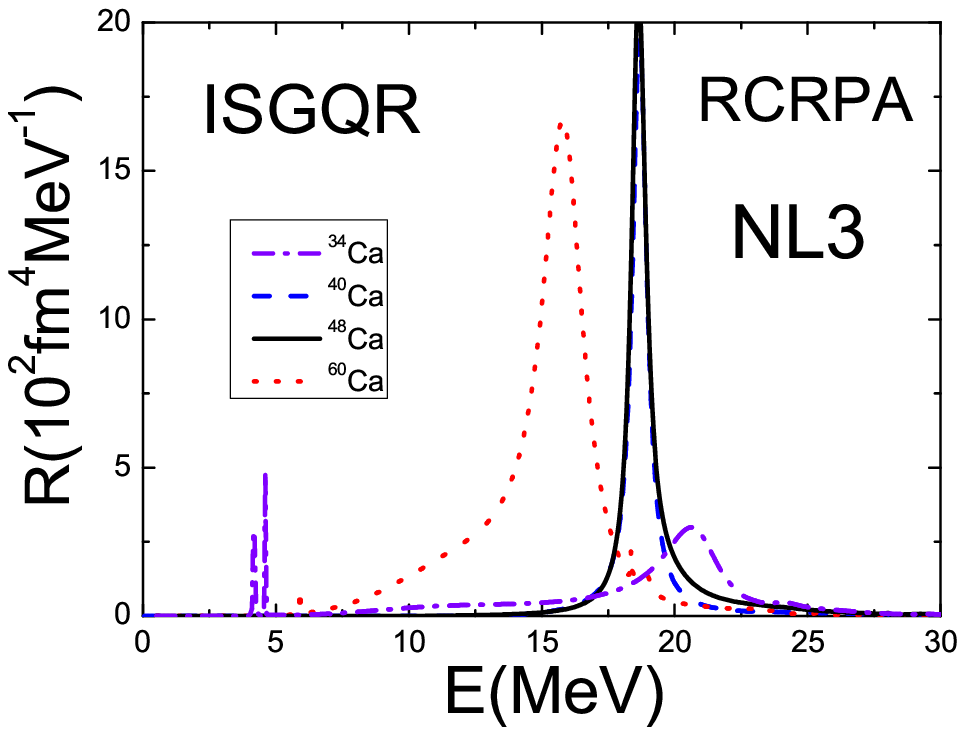}}
   \resizebox{20pc}{!}{\includegraphics{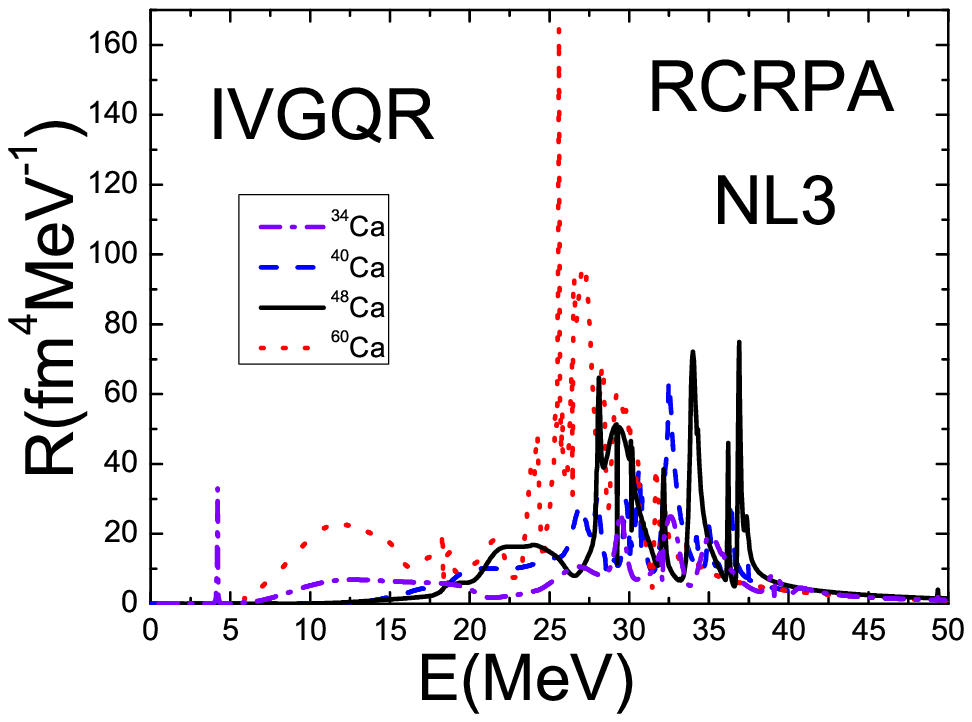}}
  \end{center}\vglue -1cm
\caption{(color online) The response   functions   of $^{34}$Ca,
$^{40}$Ca,$^{48}$Ca and $^{60}$Ca in RCRPA .The
dash-dotted,dashed,solid and dotted  curves represent the results of
$^{34}$Ca, $^{40}$Ca,$^{48}$Ca and $^{60}$Ca, respectively.}
\label{ca34ca40ca48ca60}
\end{figure}

\begin{figure}[t]
\begin{center}
   \resizebox{18pc}{!}{\includegraphics{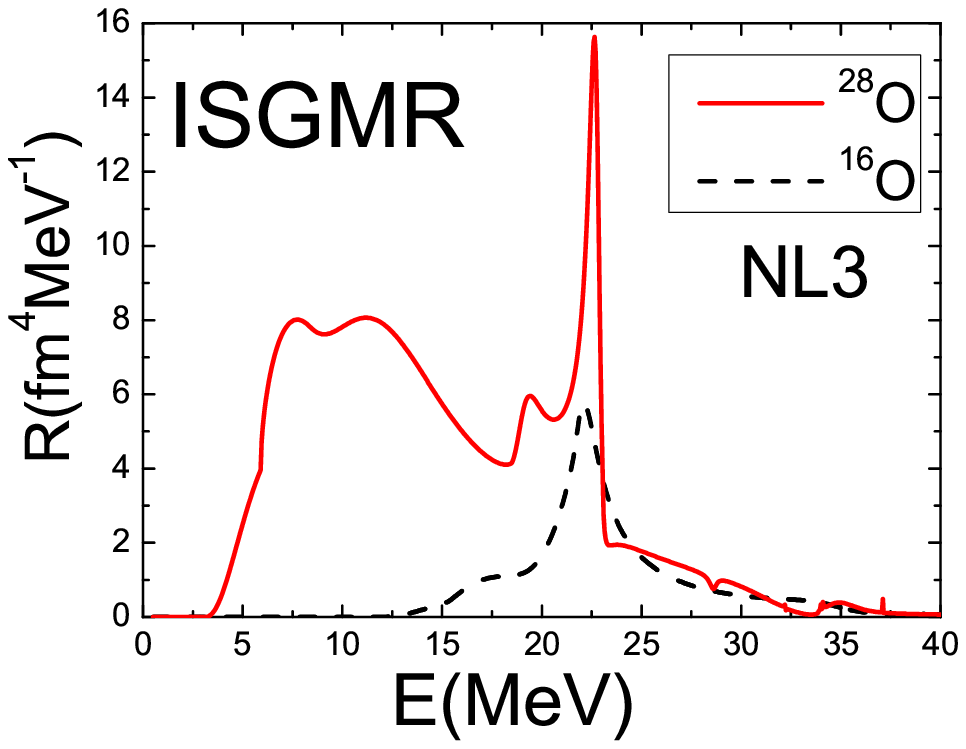}}
   \resizebox{18pc}{!}{\includegraphics{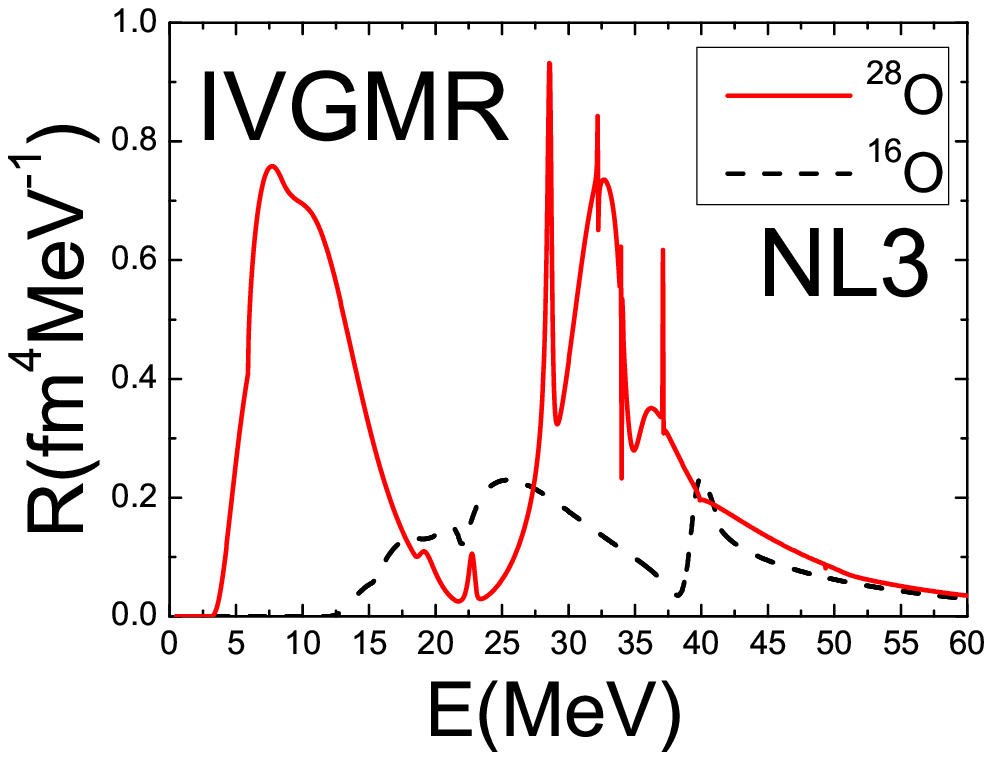}}
  \resizebox{18pc}{!}{\includegraphics{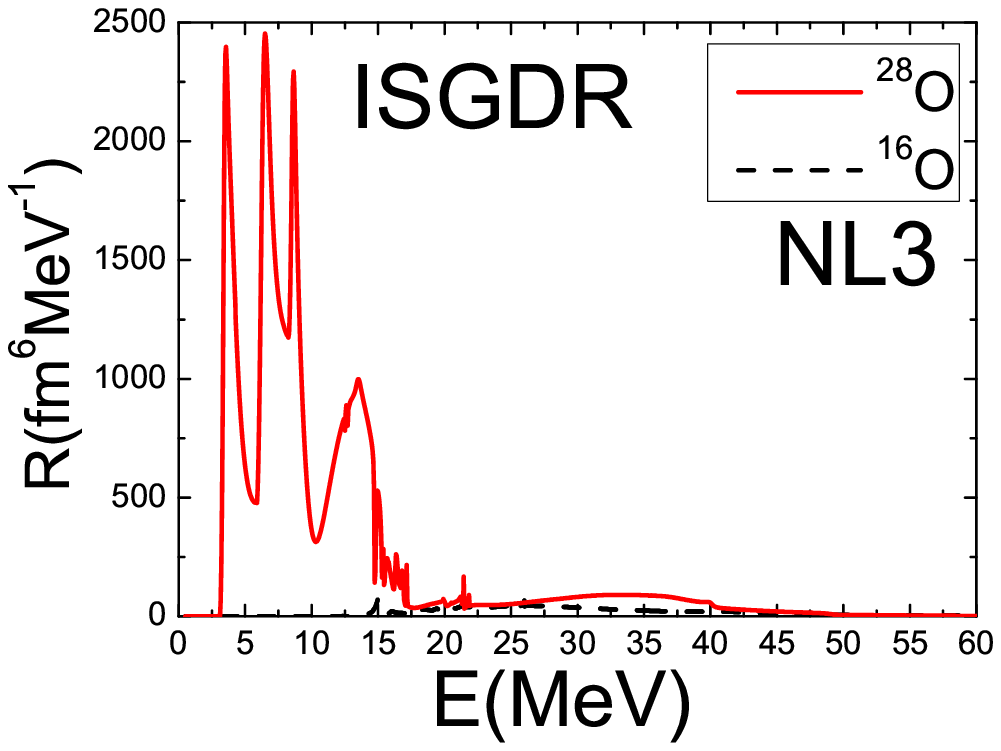}}
 \resizebox{18pc}{!}{\includegraphics{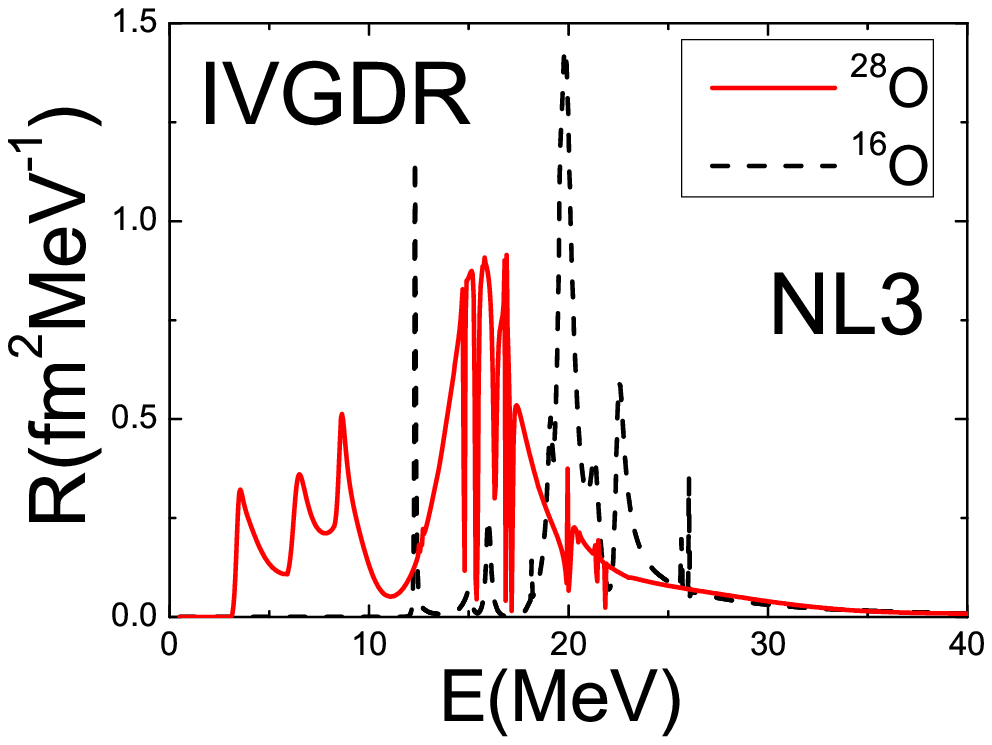}}
    \resizebox{18pc}{!}{\includegraphics{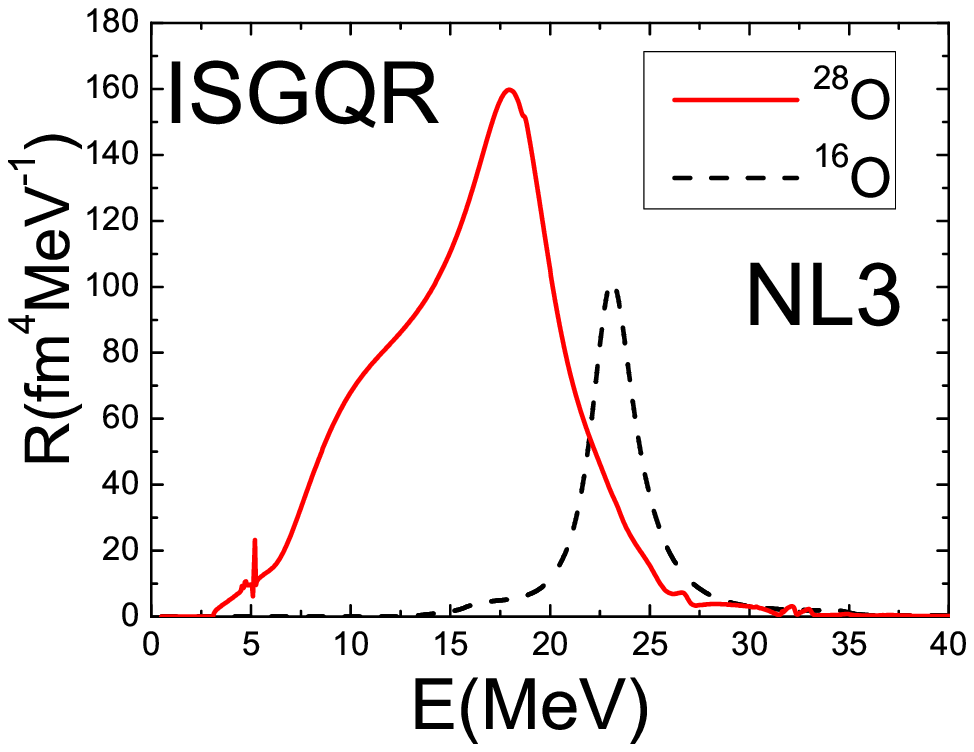}}
  \resizebox{18pc}{!}{\includegraphics{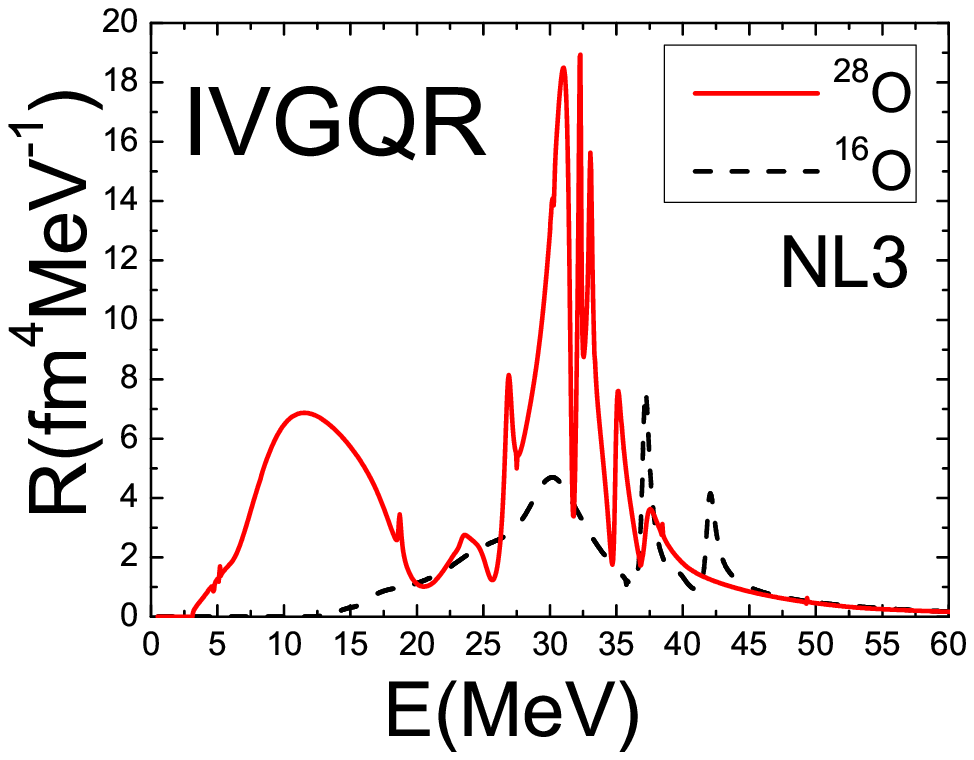}}
 \resizebox{18pc}{!}{\includegraphics{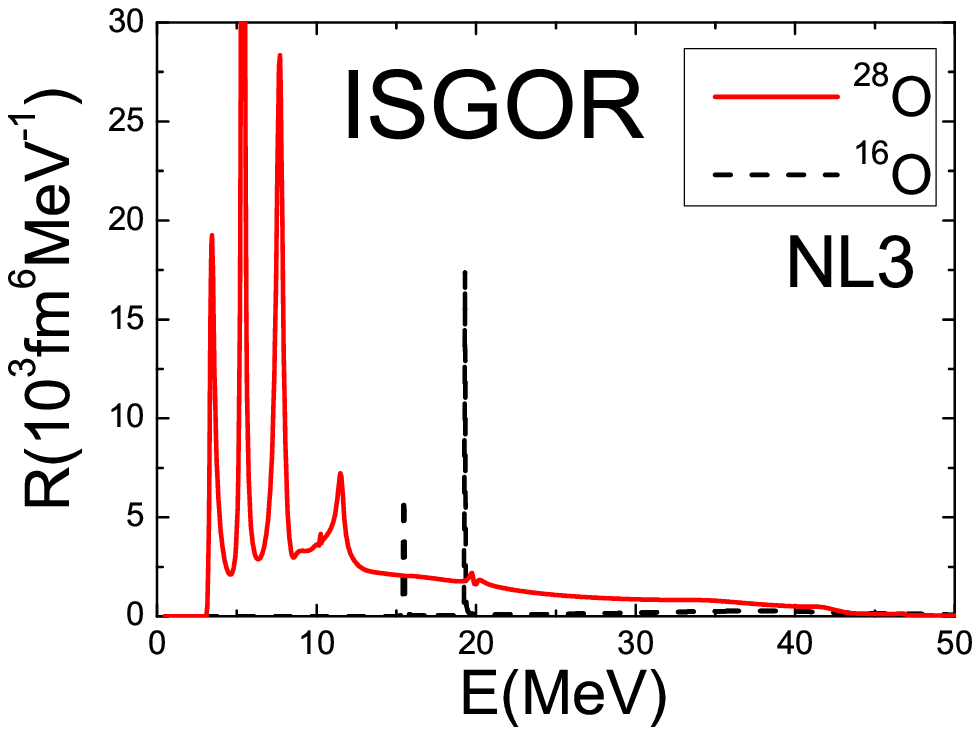}}
  \end{center}\vglue -1cm
\caption{(color online) The response   functions   of $^{16}$O and
$^{28}$O in RCRPA .The solid and dashed curves represent the results
of $^{28}$O and $^{16}$O, respectively.} \label{o16o28rcrpa}
\end{figure}

\begin{figure}[t]
\begin{center}
     \resizebox{18pc}{!}{\includegraphics{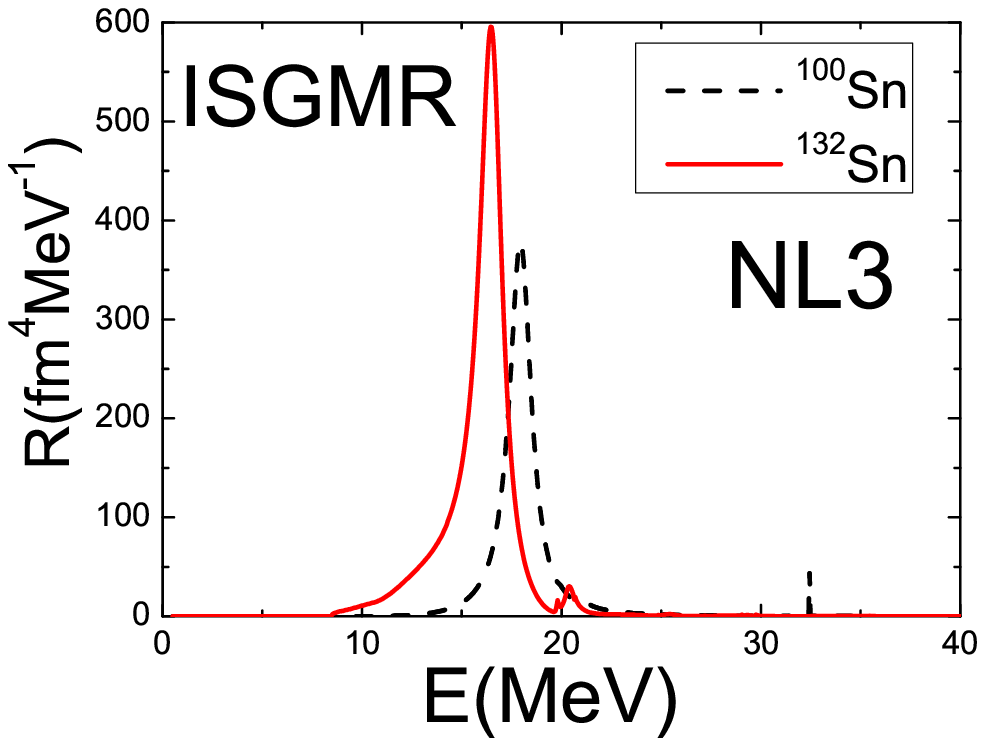}}
   \resizebox{18pc}{!}{\includegraphics{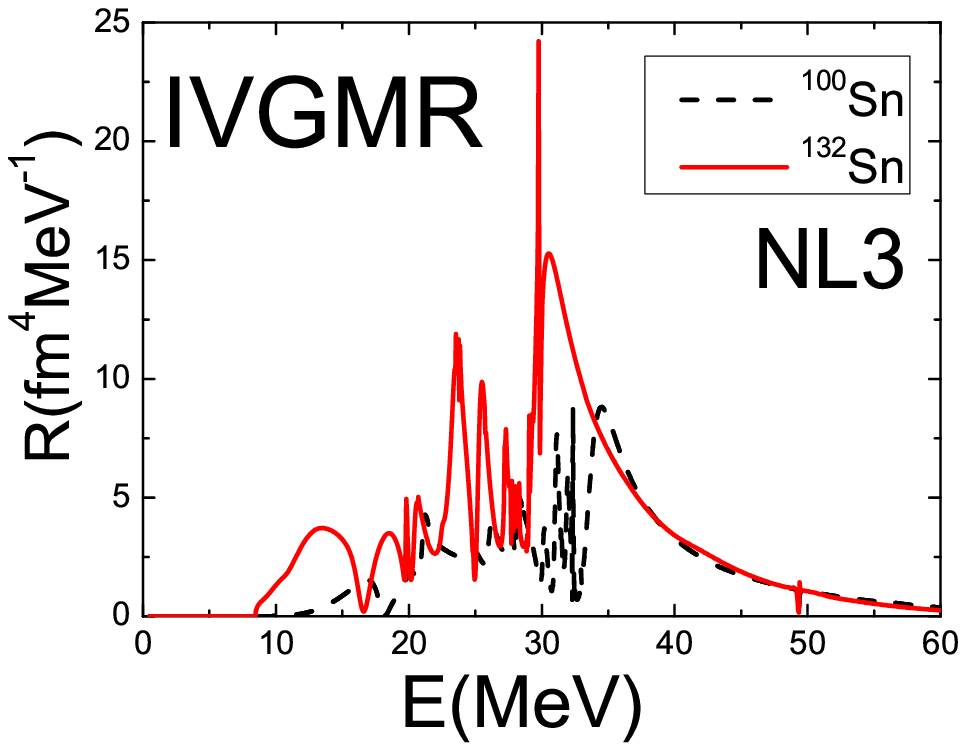}}
  \resizebox{18pc}{!}{\includegraphics{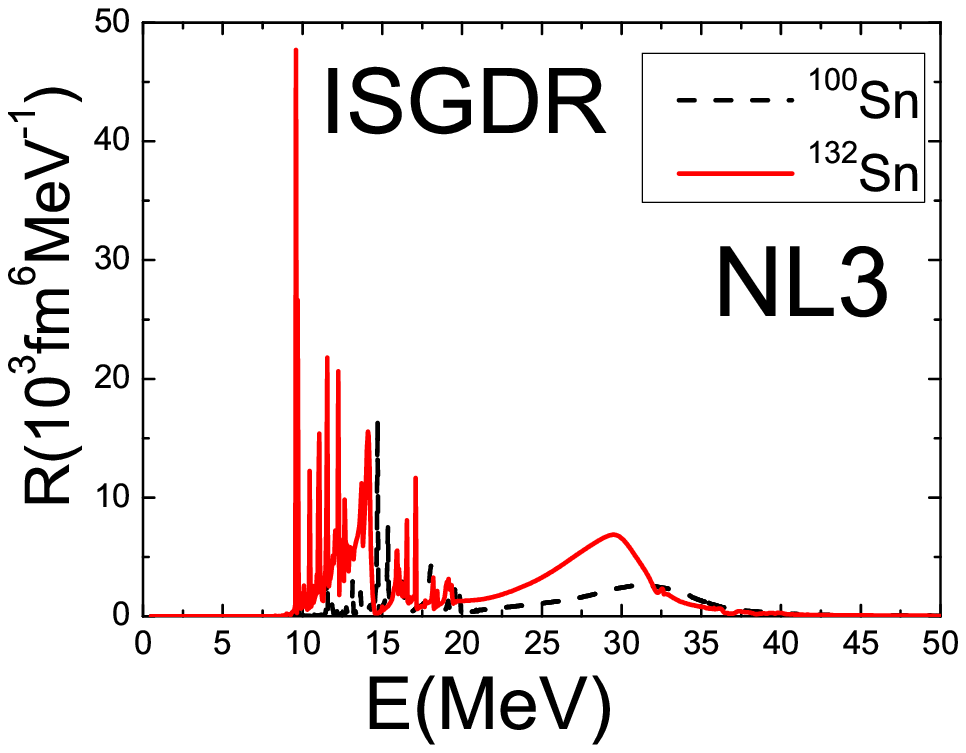}}
 \resizebox{18pc}{!}{\includegraphics{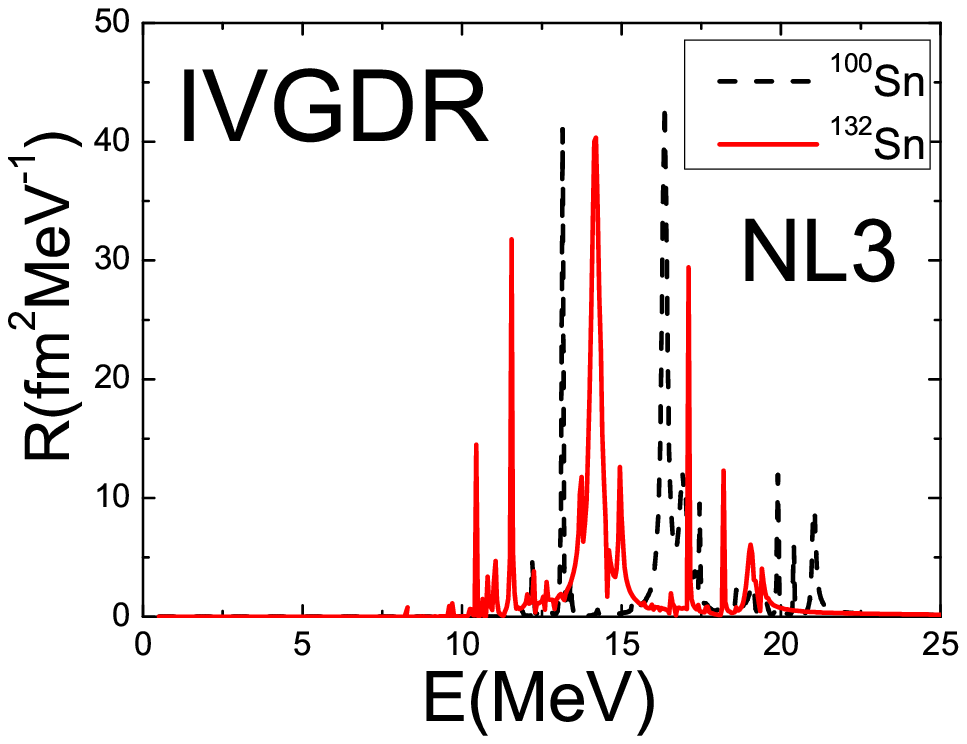}}
    \resizebox{18pc}{!}{\includegraphics{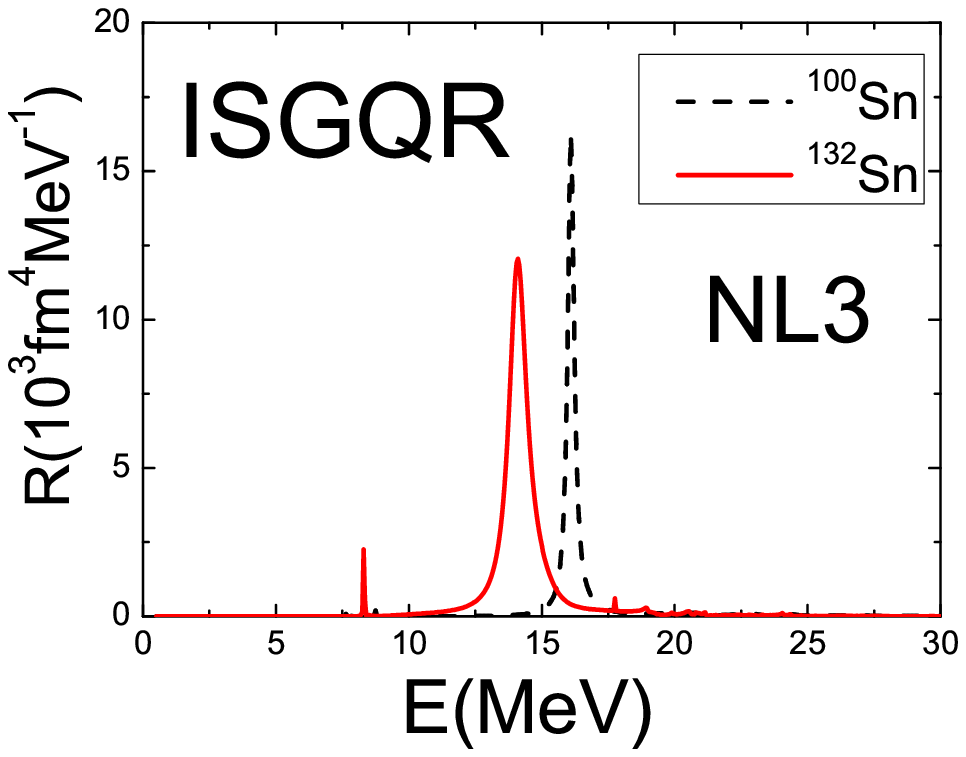}}
  \resizebox{18pc}{!}{\includegraphics{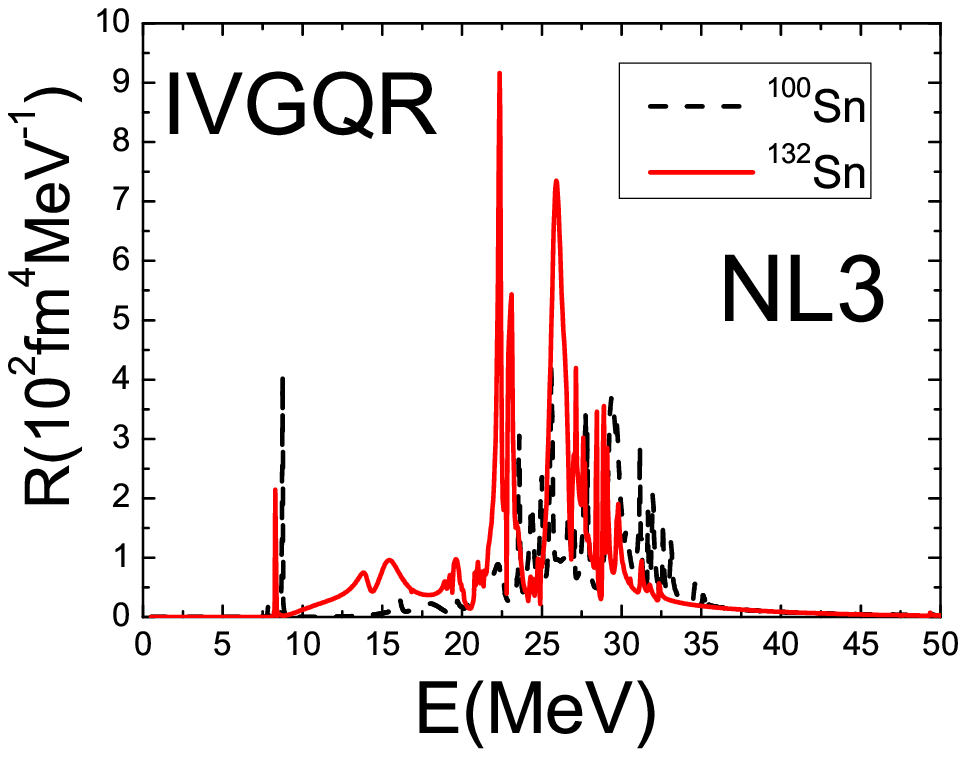}}
 \resizebox{18pc}{!}{\includegraphics{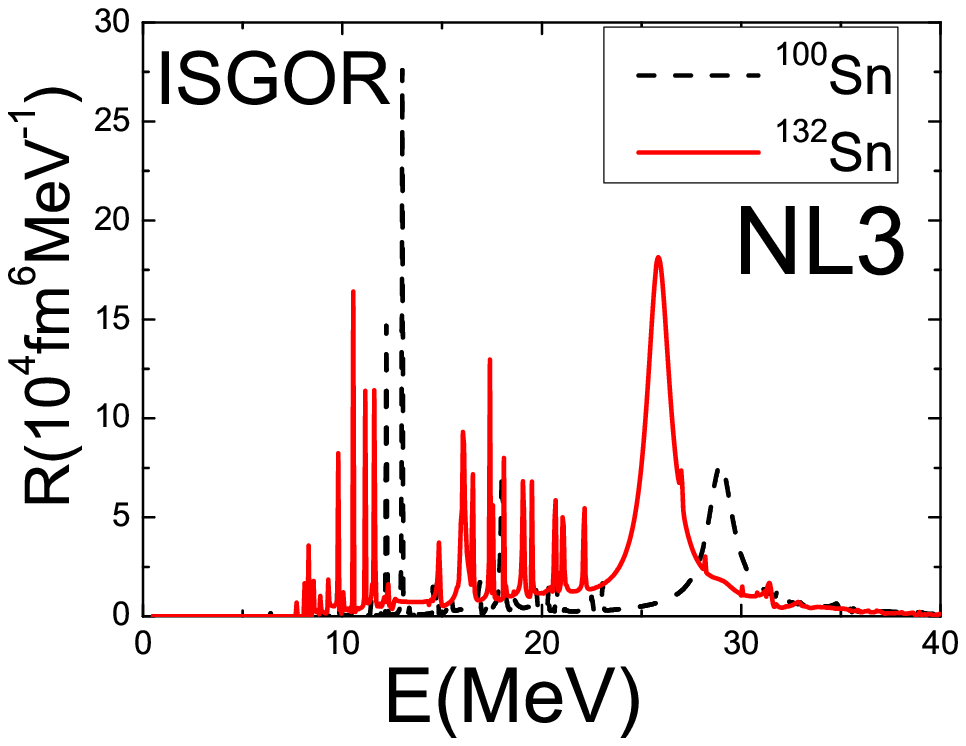}}
  \end{center}\vglue -1cm
\caption{(color online) The response   functions   of $^{100}$Sn and
$^{132}$Sn in RCRPA .The solid and dashed curves represent the
results of $^{132}$Sn and $^{100}$Sn, respectively.}
\label{sn100sn132rcrpa}
\end{figure}

\begin{figure}[t]
\begin{center}
   \resizebox{20pc}{!}{\includegraphics{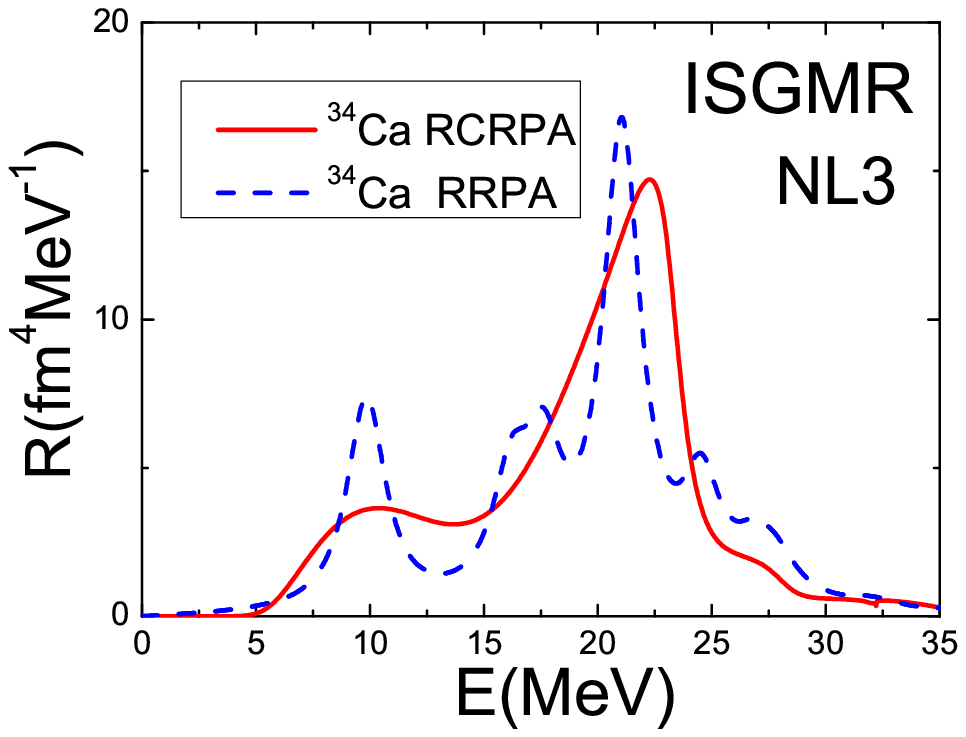}}
   \resizebox{20pc}{!}{\includegraphics{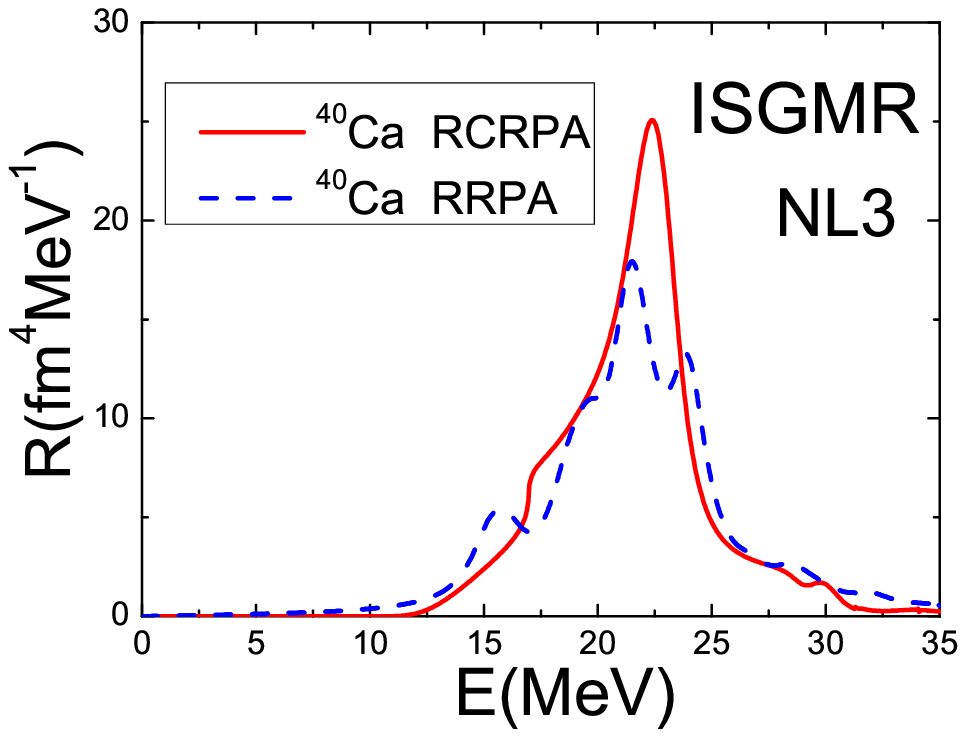}}
  \resizebox{20pc}{!}{\includegraphics{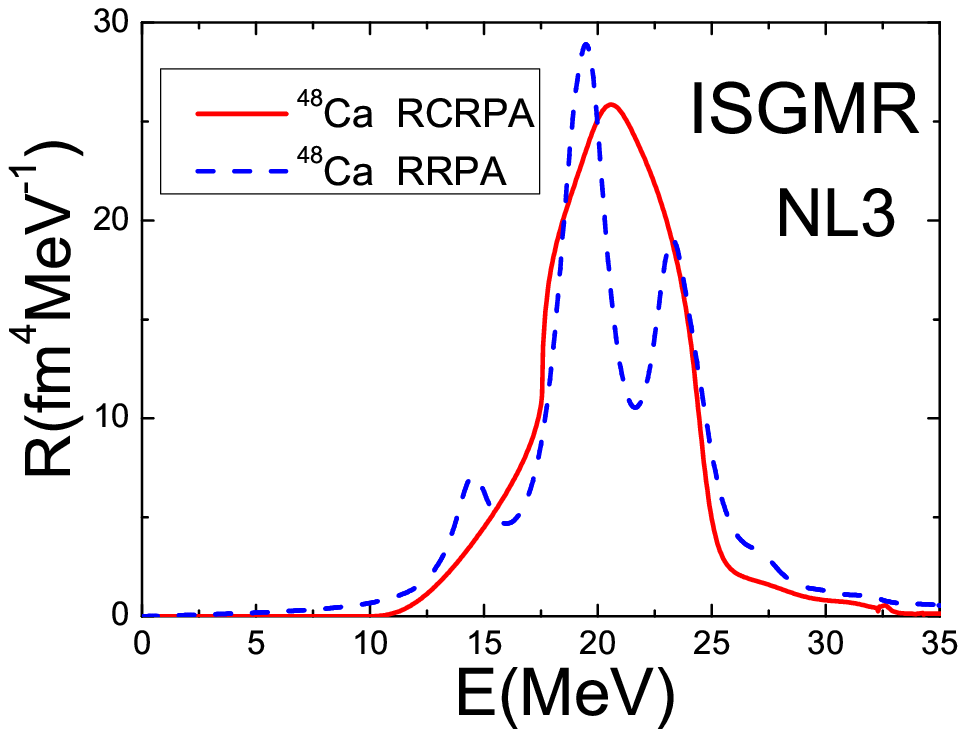}}
 \resizebox{20pc}{!}{\includegraphics{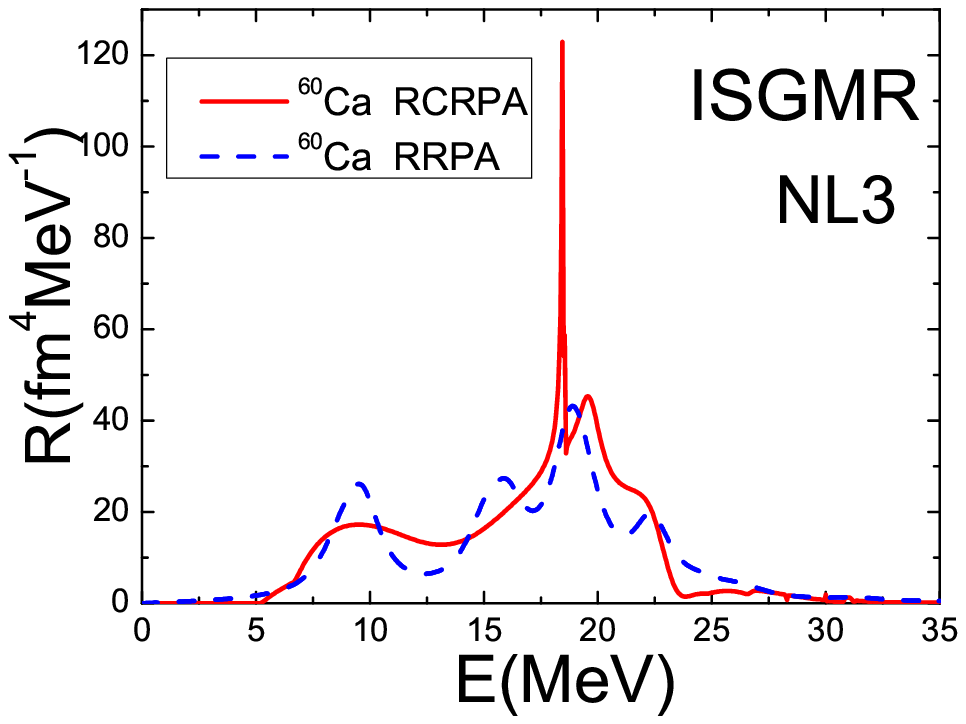}}
%   \resizebox{20pc}{!}{\includegraphics{ca34ca40ca48ca60isgqr}}
 %  \resizebox{20pc}{!}{\includegraphics{ca34ca40ca48ca60ivgqr}}
  \end{center}\vglue -1cm
\caption{(color online) The ISGMR response   functions   of
$^{34}$Ca, $^{40}$Ca,$^{48}$Ca and $^{60}$Ca in RCRPA and RRPA.The
solid and dashed  curves represent the results of RCRPA and RRPA,
respectively.} \label{ca34ca40ca48ca60-rcrpa-rrpa}
\end{figure}

\section{Collective Multipole Excitations of Exotic Nuclei  }

The response functions of isoscalar and isovector giant resonances
of multipolarities $L$ = 0-2 for $^{34}$Ca, $^{40}$Ca, $^{48}$Ca and
$^{60}$Ca calculated in RCRPA are plotted in
Fig.\ref{ca34ca40ca48ca60}. The dashed-dotted, dashed, solid and
dotted curves represent the results of $^{34}$Ca, $^{40}$Ca,
$^{48}$Ca and $^{60}$Ca, respectively. As we know, $^{60}$Ca is
neutron-rich nucleus and $^{34}$Ca is proton-rich nucleus. In
neutron-rich and proton-rich nuclei, the neutron or proton density
has different profile. From Fig.\ref{ca34ca40ca48ca60}, we find that
the neutron or proton excess has large effects on the energy
distribution of the strength, it leads to strong low-energy
excitations and pushes the centroid of the strength function to
lower energies. In the isovector resonances, one observes some
fragmentation in neutron-rich nucleus $^{60}$Ca , namely, the
neutron excess increases the fragmentation of the strength
distribution. In contrast to response function of stable nucleus
$^{48}$Ca, in neutron-rich nucleus $^{60}$Ca and proton-rich nucleus
$^{34}$Ca, strong low-energy excitations are found, especially in
the case of $^{60}$Ca. In the case of the isoscalar giant quadrupole
resonance(ISGQR), it is clearly found that the response functions of
the normal nuclei $^{40}$Ca and $^{48}$Ca are similar, in contrast,
the response function of neutron-rich nucleus $^{60}$Ca is shifted
to lower-energy region remarkablely, the response function of
proton-rich nucleus $^{34}$Ca is separated into higher-energy region
and low-energy excitations. In the compressional isoscalar giant
monopole resonance(ISGMR) and isoscalar giant dipole
resonance(ISGDR), a strong %concentration of the
strength at lower energy is found in neutron-rich nucleus $^{60}$Ca,
lower-energy excitation can also be found in proton-rich nucleus
$^{34}$Ca, but it is smaller than that of $^{60}$Ca. On the other
hand, the higher-energy part of the strength in $^{60}$Ca goes up,
while the higher-energy part in $^{34}$Ca drops down.

In addition, in order to show the above discussion is more general, %the universality of the above results,
we present the response functions of $^{16}$O and $^{28}$O for L=0-3
calculated in RCRPA in Fig.\ref{o16o28rcrpa}. The solid and dashed
curves represent the results of $^{28}$O and $^{16}$O, respectively.
We also show the response functions of $^{100}$Sn and $^{132}$Sn for
L=0-3 calculated in RCRPA
 in Fig.\ref{sn100sn132rcrpa}. The solid and dashed curves
represent the results of $^{132}$Sn and $^{100}$Sn, respectively.
Here, $^{16}$O is a normal nucleus, and $^{100}$Sn is a proton-rich
nucleus, while $^{28}$O and $^{132}$Sn are neutron-rich nuclei. From
Fig.\ref{o16o28rcrpa}-\ref{sn100sn132rcrpa}, it is also found that
the neutron excess leads to strong low-energy excitations and
increases the fragmentation of the strength distribution. It can be
seen easily that the above results are similar to those of Ca
isotopes.

It is %well
known that the continuum in the RRPA calculations is usually
discretized by the box boundary conditions, while the continuum
plays important role in description of the properties
of neutron-rich nuclei and proton-rich nuclei %becomes important
since the Fermi surface of those nuclei is close to the particle
continuum, one shall treat the
contribution of the continuum properly, %the coupling between the
%bound states and
As a result, it is required to consider the contribution of the
continuum rigorously in exotic nuclei both for the ground states and
excited states calculations. Finally, in order to clarify the effect
of treating the contribution of continuum exactly, the ISGMR
response functions given by RCRPA are compared with those obtained
by RRPA. The ISGMR response functions of $^{34}$Ca, $^{40}$Ca,
$^{48}$Ca and $^{60}$Ca given by RCRPA and RRPA are plotted in
Fig.\ref{ca34ca40ca48ca60-rcrpa-rrpa}. The solid and dashed curves
represent the results of RCRPA and RRPA, respectively. From the
Fig.\ref{ca34ca40ca48ca60-rcrpa-rrpa}, it can be noted that the
response functions calculated in RCRPA and RRPA are different from
each other, in the case of RRPA results the width of strength
functions is given artifically, in the present calculations we set
it equals to 2 MeV, while it is given automatically by the RCRPA
calculations, not only the Landau width but also the escaping width.
For example, there is a sharp peak in the ISGMR strength of
$^{60}$Ca given by RCRPA calculation, this is mainly due to the
contribution
of the single-particle resonances states in the continuum, but it can not be reproduced by the RRPA calculation.     %similar in the
%case of normal nuclei $^{40}$Ca and $^{48}$Ca, while the response
%functions are different in neutron-rich nucleus $^{60}$Ca and
%proton-rich nucleus $^{34}$Ca. It indicates that the coupling
%between the bound states and the continuum is not important in
%normal nuclei and it becomes important in neutron-rich nuclei and
%proton-rich nuclei.

\section{Summary}

In conclusion, we have studied the isoscalar and isovector
collective multipole excitations in
 %Ca isotopes
 exotic nuclei in the RCRPA framework.  The
method is based on the Green's function technique and the
contribution of the continuum spectrum  is treated exactly. We have
found strong low-energy excitations in neutron-rich nuclei and
proton-rich nuclei which is different from the case in
$\beta$-stable nuclei. Namely, the neutron or proton excess leads to
strong low-energy excitations and increases the fragmentation of the
strength distribution. The effects of treating the contribution of
continuum exactly are also discussed.% it is found that the coupling
%between the bound states and the continuum becomes important in
%neutron-rich nuclei and proton-rich nuclei.

\end{document}